\documentstyle[12pt]{article}

\newlength{\dinwidth}
\newlength{\dinmargin}
\def\lapproxeq{\lower .7ex\hbox{$\;\stackrel{\textstyle
<}{\sim}\;$}}
\def\gapproxeq{\lower .7ex\hbox{$\;\stackrel{\textstyle
>}{\sim}\;$}}
\def\be{\begin{equation}}
\def\ee{\end{equation}}
\def\bea{\begin{eqnarray}}
\def\eea{\end{eqnarray}}

\makeatletter
\def\fmslash{\@ifnextchar[{\fmsl@sh}{\fmsl@sh[0mu]}}
\def\fmsl@sh[#1]#2{%
\mathchoice
{\@fmsl@sh\displaystyle{#1}{#2}}%
{\@fmsl@sh\textstyle{#1}{#2}}%
{\@fmsl@sh\scriptstyle{#1}{#2}}%
{\@fmsl@sh\scriptscriptstyle{#1}{#2}}}
\def\@fmsl@sh#1#2#3{\m@th\ooalign{$\hfil#1\mkern#2/\hfil$\crcr$#1
#3$}}
\makeatother

\begin{document}
\titlepage
\begin{flushright}
DTP/97/112\\
December 1997 \\
\vspace*{1in}
\end{flushright}
\renewcommand{\thefootnote}{\fnsymbol{footnote}}
\begin{center}
{\Large \bf Perturbative universality in QCD jet physics} \\
\vspace*{0.7in}
Valery A.\ Khoze$^{1 \; 2}$\footnote{Talk at the XXVII International Symposium on Multiparticle Dynamics, Frascati (Rome), Italy, 8-12 September 1997.} \\
\end{center}
\vspace*{0.4in}
\begin{tabbing}
xxxxxxxx \= \kill
\> $^1$INFN - Laboratori Nazionali di Frascati, P.O. Box 13, I-00044 Frascati, Italy \\
\> $^2$Department of Physics, University of Durham, DH1 3LE, UK
\end{tabbing}

\renewcommand{\thefootnote}{\arabic{footnote}}

\vspace*{3cm}
 
\begin{abstract}
I survey some recent advances in the applications of the analytical 
perturbative approach to the description of particle distributions in 
multi-jet processes.  New tests of the perturbatively based picture in 
the (semi) soft region are discussed.
\end{abstract}

\newpage
\noindent{\large \bf  1.~Introduction} \\

These days the field of multiparticle production in QCD jets has entered a
Renaissance age.  It looks quite timely to try to realise where we are
now and where we are going.

In this talk I focus on some selected aspects of chromodynamics of jets in 
the (semi) soft region.  The main goal is to illustrate some recent impressive
phenomenological advances of the analytical perturbative approach (for reviews
see e.g. \cite{1,2,3}) which attempts to describe the gross features of the
hadronic jet-like final states without making any reference to the fragmentation
dynamics at all.  This approach is based on the so-called Modified Leading
Logarithmic Approximation (MLLA) \cite{4} and on the concept of Local Parton
Hadron Duality (LPHD) \cite{5}.

In the last years physics of hadroproduction in multi-jet events has been
very intensively studied in $e^+ e^-$, hadron-hadron and $ep$ scattering
processes.  It will certainly remain one of the main topics for studies at the
$e^+ e^-$, $pp(\overline{p})$ and $ep$ colliders of the future.  The interest in the detailed
studies of the jet chromodynamics is twofold.  On the one hand, they are 
important for testing both perturbative and non-perturbative physics of 
multiple hadroproduction, for design of experiments and the analysis of the
data.  On the other hand, the detailed knowledge of the characteristic features
of the multi-jet states could provide useful additional tools to study other
physics.  For instance, it could play a valuable role in digging out the signals
for new physics from the conventional QCD backgrounds using the colour event
portrait as a \lq \lq partonometer" mapping the basic interaction short-distance
process (for recent detailed studies and references see \cite{6}).

Nowadays, a vast amount of data from hadronic $Z^0$ decays ($\sim 20$ million 
events, about 40 identified mesons and baryons) has been accumulated in $e^+ e^-$
collisions (for reviews see e.g. \cite{3,7,8}).  New results continue to pour out
from LEP, see \cite{9}.  Recently new (very impressive) experimental data on
particle distributions in multi-jet events from HERA and TEVATRON have become
available, see e.g. \cite{10,11,12,13}.

The wealth of existing data collected in various hard processes (in hardness
interval $10-10^5 \: {\rm GeV}^2$) convincingly proves the dominant role of the 
perturbative phase of jet evolution and strongly supports the LPHD hypothesis
according to which the conversion of partons into hadrons occurs at low
virtuality scale (of order of a hadron mass), independent of the scale of the 
primary hard process, and involves only low-momentum transfer.

The LPHD allows one to relate the (sufficiently) inclusive hadronic observables
to the corresponding quantities computed for the cascading partonic system. Only
two parameters are actually involved in the perturbative description: the 
effective QCD scale $\Lambda$ and a cut-off parameter $Q_0$.  The non-perturbative
effects are practically reduced to normalization constants relating hadronic
characteristics to partonic ones.  Up to now there were no special reasons to
update the values of the free phenomenological parameters found from the first
perturbative analysis of the inclusive particle spectra in jets \cite{5}.

Rediscovery of coherence in the context of QCD in the early eighties has led to
a dramatic revision of theoretical expectations for semisoft particle
distributions.  Thus, the coherent effects in the intrajet partonic cascades, 
resulting on the average, in the angular ordering (AO) of sequential branching,
gave rise to the {\it hump-backed} shape of particle spectra --- one of the
most striking perturbative predictions \cite{5,14,15}.  It is not the softest
particles, but those with the intermediate energies $(E_h \sim E^{0.3-0.4})$
which multiply most effectively in QCD cascades. Due to the interjet coherence
which is responsible for the string \cite{16}/ drag \cite{17} effect in the
multi-jet hadronic events, a very important physical phenomenon can be 
experimentally verified, namely, the fact that it is the dynamics of the colour
which governs the production of hadrons in accordance with the QCD 
\lq \lq radiophysics" of particle flows.  Recently the first (quite impressive)
data on interjet coherence effects in $W$ + jet production from D0 \cite{18} 
have become available.

The experimental studies of the structure of the multi-jet events nicely 
demonstrate that the bright colour interference effects survive the hadronization
stage and are clearly seen in the data.  This could be taken as a strong
argument in favour of the LPHD concept.  However, despite all its
phenomenological successes, the LPHD is, by no means, a complete theoretical
scheme but rather the simplest model-independent approach.  Without doubt, 
the hadronization effects could and should be of importance in many cases.  After
all, we observe jets of hadrons in the detectors, not the quarks and gluons we
are dealing with in our perturbative calculations.  However, the dynamics of
hadronization is still not well understood from first principles and one has
to rely on the predictions of the phenomenological models, which are far from
perfect, see e.g. \cite{8,19,20}. Moreover, for many inclusive 
observables the LPHD concept (at least, in its milder formulation) is 
quantitatively realised within these algorithmic schemes.

It has to be emphasised that the LPHD lies at the very heart of the perturbative
approach, but at the same time this key hypothesis could be considered as its
Achilles heel.  One may expect that LPHD works better and better with increasing
energy since the sensitivity to the cut-off should decrease.  It seems to be quite
a delicate question of where exactly to draw the line of what precisely 
perturbative picture is capable to predict at current energies and what not.  To
find out such lines is a challenge to experiment.  For instance, one may be
tempted to ask an instructive question of what is the largest value of the
cut-off $Q_0$ which is allowed by the whole wealth of the present data (inclusive
particle spectra and correlations, multiplicity distributions, distributions of
event-shape variables, string/drag effect etc).  Certainly, the $Q_0$ scale 
definition depends on the adopted hadronization model.  Thus, for instance, 
within the Lund string scheme \cite{16}, the $Q_0$ scale above $2 {\rm GeV}$ is
disfavoured by the existing data \cite{21}.

I would expect that the allowable $Q_0$ value could be pushed down towards the
hadronic mass scale if one performs the detailed analysis of the data on the
dependence of the string/drag effect in $q \overline{q} g$ events on the particle mass
$m_h$ and $p_h^{{\rm out}}$ (momentum out of the event plane).  In my view this
may be an interesting exercise for the QCD fitting experts. \\

\newpage 

\noindent{\large \bf 2.~On inclusive particle spectra in QCD jets} \\

One of the well known (but still quite impressive) predictions of the 
perturbative scenario is the {\it hump-backed} shape of the inclusive particle
distributions in the variable $\xi = {\rm log} \; \frac{1}{x}$ with
$x = \frac{2 E_h}{\sqrt{s}}$.  At the moment all observed inclusive energy
spectra prove to be in surprisingly good agreement with the predicted by the
MLLA-LPHD approximately Gaussian-shape distribution. Moreover, the data 
collected in various hard scattering processes ($e^+ e^-$, DIS, $\overline{p} p$)
clearly demonstrate a remarkable universality of particle spectra assuming the
proper (MLLA-based) choice of the cascading evolution variables, equivalent to
the $e^+ e^- cms$ energy $\sqrt{s}$.  Recall that within the QCD cascading 
picture the evolution parameter corresponding to the struck quark jet in DIS in
Breit frame is the four-momentum transfer $\sqrt{Q^2}$ (see e.g. \cite{22,23}).
The proper energy scale for inclusive particle distribution in jets within
restricted cone $\theta_0$ measured by the CDF \cite{13} is $E$-jet $\theta_0$, 
see refs.\ \cite{2,24}.

The experimental analysis of the current jet hemisphere in DIS is the Breit
frame \cite{11} shows that the charged hadron spectrum not only has the same
shape as that seen in a single hemisphere of an $e^+ e^-$ event but also that
this shape evolves in $Q^2$ in the same way as the latter does in terms of the
$e^+ e^-$ centre-of-mass energy $\sqrt{s}$.  The measured area, peak position
and the width $\sigma$ of the spectrum confirm that the evolution variable, 
equivalent to the $e^+ e^- cms$ energy $\sqrt{s}$, is in the Breit frame $Q$.
The variation of the peak position $\xi_p^*$ with $Q$ follows the $e^+ e^-$
curve very closely.

A striking confirmation of the perturbative picture has been found by the CDF
\cite{13}.  The studies were performed of the inclusive charged particle 
momentum distributions for a variety of dijet masses $(83 < M_{jj} < 625 {\rm GeV}$)
and opening angles $\theta_0$.  The shapes of the measured $\xi$-distributions
at various $E$-jet $\theta_0$ values turn out to be remarkably close to the
MLLA expectations.  As $E$-jet $\theta_0$ increases, the peak of the spectrum,
$\xi^*$ shifts towards larger values of $\xi$ in perfect agreement with the 
MLLA predictions and $e^+ e^-$ data.

Quite challenging looks the low momentum wing of the particle spectra
$(p_h \lapproxeq 1 {\rm GeV})$ where the non-perturbative dynamics could
wash out the perturbatively based expectations.  An attempt to stretch the
perturbative predictions to the limit of their applicability (or better to say,
beyond it) has been performed in \cite{25,26}.  In particular, it was shown
that (after the proper modifications) the perturbatively based formulae allow a
sufficiently smooth transition into the soft momentum domain.  These modifications
are closely related to the colour coherence in the parton branchings and to the
space-time picture of hadroproduction in QCD jets.  Let us recall that the gluons
of long wave length are emitted by the total colour current which is independent
of the internal structure of the jet and is conserved when the partons split.
Applying the LPHD hypothesis one then expects that the hadron spectrum at very
low momenta $p$ should be nearly independent of the jet energy \cite{5,25}.  

As discussed in \cite{25}, the low momentum data could be considered as a 
further confirmation of the basic ideas of QCD coherence and LPHD.  Quantitatively,
the analysis was performed in terms of the invariant particle density 
$E \frac{dn}{d^3 p}$ for $e^+ e^-$ annihilation into hadrons at low momenta in
quite a wide $cms$ energy region (from ADONE to LEP-2).  The spectra were found
to be in a good agreement with the scaling behaviour and with analytical results.
Furthermore, the new H1 data \cite{11} are in a good agreement with the
perturbative expectations, thus confirming the universality of soft particle
production \cite{26}.

We briefly discuss here some selected issues on the inclusive  one-particle 
distributions in jets which were the starting point for the first quantitative
tests of the MLLA predictions, see e.g. \cite{2,3,5}.

Recall that within the MLLA the parton energy spectrum appears as a solution
of the corresponding Evolution Equation \cite{2,4}.  This solution can be 
presented analytically in terms of confluent hypergeometric functions depending
on two parameters, the effective QCD scale $\Lambda$ and the $k_\perp$ cut-off
$Q_0$ in the partonic cascades.  When $Q_0 = \Lambda$ the analytical result
simplifies drastically and one arrives at the so-called limiting spectrum
\cite{5} which proves to be so successful in fitting the data on charged
particle and pion production in QCD jets.

For the case of $e^+ e^-$ collisions the inclusive hadron spectrum is the sum 
of two $q$-jet distributions.  In terms of the limiting spectrum one obtains
\be
\label{eq:1}
\frac{1}{\sigma} \; \frac{d \sigma^h}{d \xi} \; = \; 2 K^h \: D_q^{{\rm lim}}
(\xi, Y)
\ee
where $K^h$ is the hadronization constant, $\sqrt{s}$ the total $cms$ energy
and $Y = {\rm log} (\sqrt{s}/2 Q_0)$. The limiting spectrum is readily 
given using an integral representation for the confluent hypergeometric 
function \cite{4,27}
\bea
\label{eq:2}
&& D_q^{{\rm lim}}  (\xi, Y) \;  =  \;  \frac{4C_F}{b} \; \Gamma (B)  \\ 
&& \times \; \int_{-\frac{\pi}{2}}^{\frac{\pi}{2}}  \; \frac{d \ell}{\pi} e^{-B\alpha} \;
\left[\frac{{\rm cosh} \alpha + (1 - 2 \zeta) {\rm sinh} \alpha}{\frac{4N_C}{b} 
\: Y \: \frac{\alpha}{{\rm sinh} \alpha}}  \right]^{B/2}  
\; \times \; I_B  \left( \sqrt{\frac{16 N_C}{b} \; Y \frac{\alpha}{{\rm sinh} 
\alpha} [{\rm cosh} \alpha + (1 - 2 \zeta) {\rm sinh} \; \alpha]} \right).  \nonumber
\eea
Here $\alpha = \alpha_0 + i \ell$ and $\alpha_0$ is determined by tanh
$\alpha_0 = 2 \zeta - 1$ with $\zeta = 1 - \frac{\xi}{Y}$. $I_B$ is the 
modified Bessel function of order $B$, where $B = a/b, \; a = 11 N_C/3
+2 n_f/3 N_C^2$, $b = (11 N_C - 2 n_f) / 3$, with $n_f$ the number of flavours
and $C_F = (N_C^2 - 1) / 2 N_C = 4/3$.  

The analysis of charged particle spectra using this distribution (e.g. \cite{3})
yields values for the effective scale parameter $\Lambda \equiv \Lambda_{ch}
\simeq 250 {\rm MeV}$.  If both parameters $Q_0$ and $\Lambda$ are kept free
in the fit one recovers the limiting spectrum with $Q_0 = \Lambda$ as best
solution.

It proves to be very convenient (see e.g. \cite{3,27}) to analyse inclusive
particle spectra in terms of the normalised moments
\be
\label{eq:3}
\xi_q \; \equiv \langle \; \xi^q  \; \rangle =  \frac{1}{\bar{{\cal N}}_E} \; \int \:
d \xi \xi^q D (\xi)
\ee
where $\bar{{\cal N}}_E$ is the multiplicity in the jet, the integral of the 
spectrum. These moments characterize the shape of the distribution and are 
independent of normalisation uncertainties.  The theoretical predictions for the
moments from the limiting spectrum are determined by only one free parameter
$\Lambda_{ch}$.  Also one defines the cumulant moments $K_q (Y, \lambda)$ or 
the reduced cumulants $k_q \equiv K_q / \sigma^q$, which are related by
\bea
\label{eq:4}
K_1 & \equiv & \overline{\xi} \; \equiv \; \xi_1 \nonumber \\
K_2 & \equiv & \sigma^2 \; = \; \langle \; (\xi - \overline{\xi})^2 \; \rangle,
\nonumber \\
K_3 & \equiv & s \sigma^3 \; = \; \langle \; (\xi - \overline{\xi})^3 \;
\rangle, \nonumber \\
K_4 & \equiv & k \sigma^4 \; = \; \langle \; (\xi - \overline{\xi})^4 \; 
\rangle \; -3 \sigma^4
\eea
where the third and fourth reduced cumulant moments are the skewness $s$ and
the kurtosis $k$ of the distribution.  If the higher-order cumulants
$(q > 2)$ are sufficiently small, one can reconstruct the $\xi$-distribution
from the distorted Gaussian formula, see \cite{27,28}. 

The cumulant moments can be obtained from
\be
\label{eq:5}
K_q (Y, \lambda) \; = \; \int_0^Y \; dy \: \left(- \frac{\partial}{\partial
\omega} \right)^q \; \gamma_\omega (\alpha_S (y)) \mid_{\omega = 0} 
\ee
where $\gamma_\omega (\alpha_S (y))$ denotes the anomalous dimension which
governs the energy evolution of the Laplace transform $D_\omega (Y)$ of the
$\xi$-distribution $D (\xi, Y)$.  In ref.\ \cite{27} the technique was developed
which allows one to derive the analytical expressions for $\frac{\langle \xi^q
\rangle}{Y^q}$.

It is worthwhile to emphasize that the basic MLLA formulae have been derived
formally in a high energy approximation.  However, even at moderate energies
$\sqrt{s}$ they are expected to give reasonable quantitative predictions
because they correspond to the exact solution of the MLLA Evolution Equation which
accounts for the main physical ingredients of parton multiplication, namely, 
colour coherence and energy balance in 2-particle QCD branching, and takes into
account also the boundary conditions for low virtuality $E \theta$. And this 
expectation proves to be well established experimentally.

As a consequence of colour coherence soft parton multiplication is suppressed, 
and the $\xi$-distribution has the form of a hump-backed plateau which is
asymptotically Gaussian in the variable $\xi$ around the maximum.  As was
mentioned before, the hump-backed plateau is among the fundamental predictions
of perturbative QCD.  Its experimental observation was welcomed by the QCD 
community but without special excitement.  Nobody nowadays expects miracles 
from the results of perturbative calculations.  However, better salesmen might 
be tempted to claim that the spectacular experimental confirmation of the 
hump-backed plateau in particle spectra has already clearly revealed the 
drastic low $x$-driven violation of the traditional DGLAP expectation 
\cite{29}, a phenomenon which many people in the other experimental 
environments (e.g. structure functions in DIS) are still so desperately 
aiming for, see e.g. \cite{10}.

A characteristic property of the limiting spectrum (\ref{eq:2}) is that it
approaches a universal finite limit at the phase space boundary $\xi = Y$ \cite{30}.
\be
\label{eq:6}
D_q^{{\rm lim}} \: (Y, Y) \; = \; C_q^g D_g^{{\rm lim}} \: (Y, Y) \; = \; C_q^g
L,
\ee
with
\be
\label{eq:7}
C_q^g \; = \; \frac{C_F}{N_C} \; = \; \frac{4}{9},
\ee
\be
\label{eq:8}
L \; = \; \frac{4 N_C}{a} \; = \; 1.069 (1.055) \; {\rm for} \; n_f \: = 
\: 3(5).
\ee

An easily accessible characteristic of the $\xi$-distribution is its maximum
$\xi^*$ which has been extensively studied by the experimental groups.  The
high-energy behaviour of this quantity for the limiting spectrum is predicted
\cite{27,28} as
\be
\label{eq:9}
\xi^* \; = \; Y \left[ \frac{1}{2} + \sqrt{\frac{C}{Y}} - \frac{C}{Y} \right]
\ee
with the constant term given by
$$
C \; = \; \frac{a^2}{16 N_C b} \; = \; 0.2915 (0.3513) \; {\rm for} \; n_f \;
= \; 3(5).
$$
Alternatively, one can compute the maximum $\xi^*$ from the Distorted Gaussian
approximation:
\be
\label{eq:10}
\xi^* \; = \; \overline{\xi} - \frac{1}{2} s \sigma
\ee
It appears that in the available energy range the expression (\ref{eq:9})
leads to a nearly linear dependence of $\xi^*$ on $Y$.  It is worthwhile to 
mention that in the large $N_C$ limit, when $11 N_C \gg 2 n_f$ the parameter
$C$ becomes independent on both $n_f$ and $N_C$ and approaches its asymptotical
value of $C = \frac{11}{3} \frac{1}{2^4} \simeq 0.23$.  Therefore in this limit
the effective gradient of the straight line is determined by such a fundamental
parameter of QCD as the celebrated $\frac{11}{3}$ factor (characterizing the
gluon self interaction) in the coefficient $b$.

As has already been mentioned, formula (\ref{eq:9}) describes surprisingly well
the observed evolution of the maximum of the spectra measured in $e^+ e^-$
collisions, current jet fragmentation at HERA and in the dijet events at 
TEVATRON (assuming a proper choice of the cascading variable).  The existing
experimental results on the $\xi^*$ evolution prove to be completely inconsistent
with cylindrical phase space expectations, see e.g.\ \cite{3,10,11}.

Let us make here a few comments concerning the application of the perturbative
analytical results to the identified particle distributions.  Recall that in 
the context of the LPHD logic the limiting formulae are applied for dealing
with the inclusive distributions of the \lq \lq massless" hadrons ($\pi$'s) and
for all charged particle spectra.  To approximate the distributions of 
\lq \lq massive" hadrons $(K, \rho, p \dots)$ the partonic formulae truncated at
different cut-off values $Q_0 (Q_0 (m_h) > \Lambda)$ could be used, e.g.\
\cite{2,3}.  Within the framework of the LPHD-MLLA picture there is no recipe
for relating $Q_o$ to the masses of the produced hadrons and their quantum
numbers.  One needs further detailed phenomenological studies of the $Q_0$ 
dependence of the spectra of identified particles/resonances.  Here also 
the data on different hadron species from jets at the TEVATRON and HERA would 
be very helpful.

The analytical expressions for the truncated parton distributions representing
the exact solution of the Evolution Equation \cite{2,4,5} are not transparent
for physical interpretation and are not easily suited to straightforward numerical
calculations. However, one can represent the results in terms of distorted 
Gaussian distribution for $D (\xi, Y, \lambda)$ with $\lambda = {\rm log} 
\frac{Q_0}
{\Lambda}$.  The MLLA effects are encoded in terms of the analytically 
calculated (for $Q_0 \neq \Lambda$) shape parameters \cite{27} $\overline{\xi},
\sigma, k, s,$ see eq.\ (\ref{eq:4}).  The mean parton multiplicity can be 
written \cite{5} in a compact form in terms of modified Bessel (MacDonald)
functions $I_\nu (x)$ and $K_\nu (x)$,
\bea
\label{eq:11}
N_A (Y,\Lambda) & = & C_A^g \; x_1 \; \left( \frac{z_2}{z_1} \right)^B \;
[I_{B + 1} (z_1) K_B (z_2) \nonumber \\ 
& + & K_{B + 1} (z_1) I_B (z_2)],
\eea
\be
\label{eq:12}
z_1 \; = \; \sqrt{\frac{16 N_C}{b} (Y + \lambda)}, \quad z_2 \; = \; 
\sqrt{\frac{16 N_C}{b} \lambda}
\ee
Here $A = q, \; g$ denotes the type of jet $(C_g^g = 1, C_q^g = \frac{C_F}{N_C})$. 
The first term in square brackets in (\ref{eq:11}) increases exponentially with
$\sqrt{Y}$ while the second term decreases.  Its role is to preserve the initial
condition for the jet evolution, namely, $N_g = 1$.

MLLA predicts the energy independent shift of the peak position for truncated
parton distributions as compared to the limiting spectrum \cite{27}.  The present
data on the identified particle spectra well confirm the perturbative expectation
that for different particle species the energy dependence of $\xi^*$ is universal.

The MLLA-LPHD predictions have been successfully confronted with the data on the
identified particle distributions (see e.g.\ \cite{3,7,8} and references therein).
In particular, the bell-shaped form of the spectra and their energy evolution 
are in a fairly good agreement with the perturbative predictions.

Finally, let us turn to the tests of the perturbatively based picture in the soft
region, see \cite{25,26,27}.  Without doubt, it is not a priori clear at all, 
whether one can appeal to the perturbative expertise when exploring the low
momentum domain.  However, an attempt to stretch the perturbative expectations to
the limit of their applicability looks quite intriguing.  This could, in principle,
provide a clue for understanding of some conceptual problems of the LPHD.  Whether
or not the transition between two stages of jet development is soft is a question
for experiment.

Certainly, within the perturbative framework there is no unique recipe of how to
modify the relation between parton and hadron distributions in order to enter
smoothly the soft domain, see discussion in \cite{25,26}.  Here we shall follow
an ancient route proposed in \cite{31} (see also \cite{26,27}) which is based on
the phase-space arguments.

Let us recall that at low momenta the invariant density of hadrons 
$E \frac{dn}{d^3 p}$ can be rewritten as
\be
\label{eq:13}
E \frac{dn}{d^3 p} \; \sim \; \frac{\overline{W}_1 (s, E \sqrt{s})}{s},
\ee
where $\overline{W}_i (s, E \sqrt{s})$ are the standard $e^+ e^-$ analogues of 
the DIS structure functions $W_i (q^2, \nu)$.  As well known, $\overline{W}_i 
(s, E \sqrt{s})$ are related to the matrix elements of the current commutators
and should be regular when $p \rightarrow 0$.  It is then a general requirement
that the hadronic density $E \frac{dn}{d^3 p}$ approaches a constant limit 
when $p \rightarrow 0$.  As demonstrated in Refs.\ \cite{25,26} this is well
established experimentally.  

In Ref.\ \cite{27} a simple prescription has been discussed of how to modify
(\ref{eq:1}) in order to satisfy (\ref{eq:13}) at low particle momenta.
\be
\label{eq:14}
\frac{1}{\sigma} \; \frac{d \sigma^h}{d \; {\rm log} \; p} \; = \; 2 K^h \left(
\frac{p}{E} \right)^3 D_q^{{\rm lim}} \; (\xi_E, Y), 
\ee
with $\xi_E = {\rm log} \frac{\sqrt{s}}{2 E}$.

With this prescription one arrives at the following expression for the invariant
hadronic density in the case of $e^+ e^-$ annihilation
\be
\label{eq:15}
E \frac{dn}{d^3 p} \; = \; 2 K^h \frac{1}{(4 \pi E^2)} \; D_q^{{\rm lim}}
(\xi_E, Y)
\ee
As it is easy to see from eqs.\ (\ref{eq:6}) and (\ref{eq:15}), hadronic density
approaches a constant limit at $p \rightarrow 0$.

For large energies $E \gg \Lambda$ Eqs.\ (\ref{eq:14}) and (\ref{eq:15})
coincide with the standard MLLA-LPHD relations.  Let us recall that the low
momentum region in charged particle spectra is dominated by pions and that the 
MLLA limiting spectrum provides a fairly good description of pion spectra at 
relativistic energies with
\be
\label{eq:16}
K^\pi \simeq 1.1 \; {\rm and} \; Q_0 \; = \; \Lambda \; \simeq \; 150 {\rm MeV},
\ee
see e.g.\ \cite{3,27}.

It is interesting to note that with these parameters the invariant pion density
at the very edge of the phase space is given by
\bea
\label{eq:17}
&& E_\pi \frac{dn}{d^3 p_\pi} \; = \; \frac{8}{9} K^\pi \frac{4N_C}{a} \frac{1}
{4 \pi Q_0^2} \; \approx \nonumber \\
&& \approx \; 0.9 \frac{1}{4 \pi m_\pi^2} \; \approx \; 4 
{\rm GeV}^{-2}
\eea

It looks quite challenging that the perturbatively based result (\ref{eq:17})
appears to be so close to the natural hadronic scale for the pion density.
Moreover, as shown in Refs.\ \cite{25,26}, the experimental data on soft pion
(and soft charged particle) production seem to favour the limiting value 
(\ref{eq:17}).  

As has already been mentioned the observed production rates of soft particles
have proven to be practically independent of the energy of parent parton. Such
scaling behaviour has been nicely demonstrated in both  $e^+ e^-$ and DIS 
interactions, over a wide range of {\it cms} energies, in which the data move toward
a common limit as the particle momenta become small, see \cite{11,25,26}. This 
could be considered as strong evidence in support of the LPHD.  Recall that the
LPHD is deeply rooted in the space-time picture of the hadroproduction in the
QCD cascades, e.g.\ \cite{2}. Thus, within it, in the process $e^+ e^- 
\rightarrow
q \overline{q}$ the first hadrons are formed at the time $t \sim t_{{\rm crit}}
\sim R \; (R \approx 1$ fm is a characteristic space-time scale of the strong
interactions) with $p \sim p_\perp \sim R^{-1} \sim \Lambda$.  It is the moment
when the distance between the outgoing $q$ and $\overline{q}$ approaches $R$. 
At $t > t_{{\rm crit}}$ the two jets are separated as globally branched, and 
the parton cascades develop inside each of them.  The gluon bremsstrahlung 
becomes intensive only when the transverse distance between any two 
colour partons exceeds $R$.  

With increasing time the partons with larger and larger energies $E \sim
\frac{t}{R^2}$ hadronize (inside-outside chain). In this picture soft
particles with $E \sim R^{-1}$ produced at the lower edge of the perturbative
phase space play a very special role.  Their production rate is practically
unaffected by the QCD cascading, and their formation is a signal of switching on
the real strong interactions $(\alpha_S \sim 0(1))$.  In some sense these 
particles can be considered as the eye-witnesses of the beginning of
the \lq \lq hadronization wave". \\

\noindent{\large \bf 3.~Colour related phenomena in multi-jet events} \\

It was realised long ago (see e.g.\ \cite{2} and references therein) that the
overall structure of particle distributions in multi-jet events in hard
scattering processes (event portrait) is governed by the underlying colour
dynamics at short distances.  The existing experimental data clearly show
in favour of interjet colour coherence, see e.g.\ \cite{3,7,8}.  Here we shall
briefly discuss some new results on QCD radiophysics of particle flows in 
multi-jet events, see also \cite{32}.  The main lesson from the recent impressive
studies is that now we have (quite successfully) entered the stage of
quantitative tests of the details of colour drag phenomena.

The interjet coherence phenomena were intensively studied at LEP1, TRISTAN and
TEVATRON.  Let us mention a few new facts concerning comparison with the 
analytical QCD predictions.

DELPHI \cite{33} has performed the first quantitative verification of the
perturbative prediction \cite{17} for the ratio $R_\gamma$
\be
\label{eq:18}
R_\gamma \; = \; \frac{N_{q \overline{q}} (q \overline{q} g)}{N_{q \overline{q}}
(q \overline{q} \gamma)}
\ee
of the particle population densities in the interquark valley in the
$e^+ e^- \rightarrow q \overline{q} g$ and $ e^+ e^- \rightarrow q \overline{q}
\gamma$ events.  For a clearer quantitative analysis a comparison was performed
for the $Y$-shaped symmetric events using the double vertex method for the
$q$-jet tagging.  The ratio $R_\gamma$ of the charged particle flows in the
$q \overline{q}$ angular interval [$35^0, \; 115^0$] was found to be 
\be
\label{eq:19}
R_\gamma^{{\rm exp}} \; = \; 0.58 \pm 0.06.
\ee
This value is in a fairly good agreement with the expectation following from
\cite{17} at $N_C = 3$, for the same angular interval
\be
\label{eq:20} 
R_\gamma^{th} \; \approx \; \frac{0.65 N_C^2 - 1}{N_C^2 - 1} \; \approx \; 0.61.
\ee
The string/drag effect is quantitatively explained by the perturbative prediction
and the above ratio does not appear to be affected by hadronization effects in
an essential way.

Another new result \cite{33} concerns the analysis of the threefold symmetric
$e^+ e^- \rightarrow q \overline{q} g$ events using the double vertex tagging
method.  It is shown that the string/drag effect is clearly present in these
fully symmetric events and it cannot be an artefact due to kinematic selections.
Quantitatively, comparing the minima located at $\pm [50^0, \: 70^0]$, the
particle population ratio $R_g \; = \; N_{q g} / N_{q \overline{q}}$ in the 
$q - g$ and $ q - \overline{q}$ valleys is measured to be
\be
\label{eq:21}
R_g^{exp} \; = \; 2.23 \pm 0.37
\ee
This is to be compared with the asymptotic prediction $R_g = 2.46$ for 
projected rates at central angles, whereas for the above angular interval one
finds \cite{3,17}
$$
R_g^{th} \; \approx \; 2.4.
$$
in good agreement with the experimental value.

If one allows for arbitrary 3-jet kinematic configurations new information can 
be obtained about the evolution of the event portrait with the variation of
event topology, see \cite{3,34}.  Recently ALEPH \cite{35} and DELPHI \cite{36}
have demonstrated that, in agreement with the QCD radiophysics \cite{2}, the 
mean event multiplicity in three jet events depends both on the jet energies and
on the angles between the jets.  These results clearly show the topological
dependence of jet properties which was predicted analytically.

Identification of charged hadrons $(\pi^\pm, K^\pm$ and $\overline{p}$) has
allowed ALEPH \cite{37} to study mass dependence of the interjet $R_g$ values.
In full agreement with the perturbative expectations \cite{17,2,3} there is no 
strong mass dependence at LEP-1 energies.

Finally, let us note that L3 and OPAL \cite{38} have studied the dependence of the
colour drag on out-of-plane momentum $p^{{\rm out}}$.  In agreement with the
predictions of \cite{17} the dependence on $p^{{\rm out}}$ was found to be 
significantly weaker than at lower energies.  Recall that the dependence of the 
magnitude of the string/drag effect on $p^{{\rm out}}$ (and registered particle
mass) has to vanish asymptotically in the perturbative approach.

Recently the D0 Collaboration has reported the first results on colour 
coherence studies in $W$ + jet events \cite{18}.  One of the instructive 
measurements concerns the ratio for soft particle production in the event
plane to the transverse plane.  This quantity proves to be insensitive to the
overall normalization of the individual distributions and to detector effects.
The experiment shows very good agreement with the perturbative expectation
\cite{6} for this ratio.

The clear observation of interjet interference effects gives another strong
evidence in favour of the perturbative picture of multihadron production.  The
collective nature of multi-parton final states reveals itself here via the QCD
wave properties of the multiplicity flows.  The detailed experimental studies 
of the colour-related effects are of particular interest for better understanding
of the dynamics of hadroproduction in the multi-jet events.  For instance, under
special conditions some subtle interjet interference effects, breaking the 
probabilistic picture, may even become dominant, see Refs.\ \cite{17,39}.  We
remind the reader that QCD radiophysics predicts both attractive and repulsive
forces between the active partons in the event.  Normally the repulsion effects
are small, but in the case of colour-suppressed $O (\frac{1}{N_C^2})$ phenomena
they may play a leading role.  It should be noted that in Refs.\ \cite{17,39}
the interjet collective effects were viewed only on a completely inclusive basis, 
when all the constituents of the multi-element colour antenna are simultaneously
active. A challenging possibility to operate within the perturbative scenario
with the complete collective picture of an individual event (at least at very
high energies) was discussed in \cite{40}.  The topologometry on the event-by-event
basis could turn out to be more informative than the results of measurements 
averaged over the events.

Recall, that there is an important difference between the perturbative 
radiophysics and the parton-shower Monte Carlo models.  The latter not only
allow but even require a completely exclusive probabilistic description.
Normally (such as in the case of $e^+ e^- \rightarrow q \overline{q} g$) the
two pictures work in a quite peaceful coexistence; the difference only
becomes drastic when one deals with the small colour-suppressed effects.

Let us emphasize that the relative smallness of the non-classical effects by 
no means diminishes their importance.  This consequence of QCD radiophysics is
a serious warning against the traditional ideas of independently evolving
partonic subsystems.  So far (despite the persistent pressure from the
theorists) no clear evidence has been found experimentally in favour of the
non-classical colour-suppressed effects in jets, and the peaceful coexistence 
between the perturbative interjet coherence and colour-topology-based 
fragmentation
models remains unbroken.  However, these days the colour suppressed 
interference
effects attract increased attention.  This is partly boosted by the findings
that the QCD interference (interconnection) between the $W^+$ and $W^-$ 
hadronic
decays could affect the $W$ mass reconstruction at LEP-2, see e.g.\ \cite{41}.

Finally, let us recall that the colour-related collective phenomena could 
become a phenomenon of large potential value as a new tool helping to 
distinguish
the new physics signals from the conventional QCD backgrounds (e.g.\ \cite{2,3,6}). \\

\noindent{\large \bf 4.~Conclusion} \\

During the last few years the experiments have collected exceedingly rich new 
information on the dynamics of hadronic jets --- the footprints of QCD partons.
New QCD physics results from LEP2, TEVATRON and HERA continue to pour out 
shedding light on various aspects  of hadroproduction in multi-jet events.  
The existing data convincingly show that the analytical  perturbative approach 
to QCD jet physics is in a remarkably healthy shape.

The key concept of this approach is that the conversion of partons into hadrons
occurs at a low virtuality scale, and that it is physics of QCD branchings which
governs the gross features of the jet development.  Thus, the perturbative 
universality of jets appearing in different hard processes is nicely confirmed.
Moreover, the data demonstrate that the transition between the perturbative
and non-perturbative stages of jet evolution is quite smooth, and that the 
bright colour coherence phenomena successfully survive the hadronization 
stage.

LEP1 proves to be a priceless source of information on the QCD dynamics.  It 
has benefited from the record statistics and the substantial lack of background.  
We have learned much interesting physics, but the need for further detailed
analyses of the data recorded at LEP1 has not decreased.

The programs of QCD studies at LEP2 and at future linear $e^+ e^-$ colliders
look quite promising.  The semisoft QCD physics becomes one of the important
topics for investigation in the TEVATRON and HERA experiments.

Concluding this talk let me emphasize once more that, of course, there is no
mystery within the perturbative QCD framework.  One is only supposed to perform
the calculational routine properly.  So it is entirely unremarkable that the
quantum mechanical interference effects should be observed in the perturbative
results.  Of real importance is that the experiment demonstrates that the
transformer between the perturbative and non-perturbative phases acts very
smoothly.  This message could (some day) shed light on the mechanism of colour
confinement.  \\

\noindent{\large \bf Acknowledgements} \\

I am grateful to A.\ De Angelis, Yu.\ Dokshitzer, S.\ Lupia, W.\ Ochs, R.\ Orava,
G.\ Rudolph, T.\ Sj\"{o}strand and W.J.\ Stirling for useful discussions.

\end{document}